\documentstyle[aps,pra]{revtex}

\input{psfig}

\newcommand{\bra}[1]{\langle #1 |\,}
\newcommand{\ket}[1]{\,| #1 \rangle}

\newcommand{\mat}[1]{\underline{\underline{#1}}}
\newcommand{\cH}{{\hat{{\cal{H}}}}}
\newcommand{\m}{\hat{\mu}}
\newcommand{\bpsi}{\bra{\Psi_o}}
\newcommand{\kpsi}{\ket{\Psi_o}}
\newcommand{\bphi}{\bra{\Phi_o}}
\newcommand{\kphi}{\ket{\Phi_o}}
\newcommand{\Self}{{\cal{S}}}
\newcommand{\beq}{\begin{equation}}
\newcommand{\eeq}{\end{equation}}
\newcommand{\beqs}{\begin{eqnarray}}
\newcommand{\eeqs}{\end{eqnarray}}
\newcommand{\zero}{{(0)}}		
\newcommand{\first}{{(1)}}		
\newcommand{\Ho}{{\hat{H}_o}}	
\newcommand{\ham}{{\hat{H}}}		
\newcommand{\calH}{{\cal H}}
\newcommand{\ky}[1]{| Y_{#1} \rangle}
\newcommand{\by}[1]{\langle Y_{#1} |}
\newcommand{\sep}{{\scriptscriptstyle FOSEP}}	
\newcommand{\rpa}{{\scriptscriptstyle RPA}}	
\newcommand{\tda}{{\scriptscriptstyle TDA}}	
\newcommand{\mSsep}{\mat{\Self}^\sep}	
\newcommand{\mMsep}{\mat{M}^\sep}	
\newcommand{\mMrpa}{\mat{M}^\rpa}	
\newcommand{\nq}{{\overline{n}}}	
\newcommand{\ph}{ph}	
\newcommand{\hp}{hp}	

\begin{document}
\draft

\title{First Order Static Excitation Potential:
Scheme for Excitation Energies and Transition Moments}

\author{Joachim Brand and Lorenz S. Cederbaum}

\address{Theoretische Chemie, Universit\"at Heidelberg, \\
	Im Neuenheimer Feld 253, D--69120 Heidelberg, Germany}

\date{October 23, 1997, {\em published in} Phys.\ Rev.\ A {\bf 57} 4311 (1998)}

\maketitle


\begin{abstract}
We present an approximation scheme for the calculation of the
principal excitation energies and transition moments of finite
many-body systems.  The scheme is derived from a first order
approximation to the self energy of a recently proposed extended
particle-hole Green's function.  A hermitian eigenvalue problem is
encountered of the same size as the well-known random phase
approximation (RPA).
 We find that it yields a size consistent description of the
excitation properties and removes an inconsistent treatment of the
ground state correlation by the RPA. By presenting a hermitian
eigenvalue problem the new scheme avoids the instabilities of the RPA
and should be well suited for large scale numerical calculations.
These and additional properties of the new approximation scheme are
illuminated by a very simple exactly solvable model.
\end{abstract}



\section*{Introduction}

Many--body Green's function theory has provided several standard
approximation schemes for the calculation of excitation properties in
atoms, molecules, and atomic nuclei.  Setting out from a convenient
single--particle description like the Hartree--Fock approximation,
these schemes usually lead to a matrix eigenvalue problem. This
supplies approximations for the excitation energies and
transition operator matrix elements of the system. The starting point is
usually the well--known {\em polarisation propagator}
\cite{fetter71}, a fundamental two--particle Green's function of
nonrelativistic many--body theory. It is given by a sum of two terms
that are related by symmetry (see e.~g.~Ref.~\cite{schirmer82}).
\beq \label{pol_prop}
  \Pi_{rs,r's'}(\omega) = \Pi^{\ph}_{rs,r's'}(\omega) +
  \Pi^{\hp}_{rs,r's'}(\omega) 
\eeq
The so--called {\em particle--hole part}
$\Pi^{\ph}_{rs,r's'}(\omega)$ already contains all the physically relevant
information exhibiting single poles in the energy variable $\omega$ at
the exact excitation energies of the system. Its name originates in
the 
single particle picture which is usually taken as the zeroth order in
a perturbation theoretical treatment. Zeroth order contributions to
$\Pi^{\ph}$ only arise if the indices $r$ and $r'$ are particle indices
and $s$ and $s'$ hole indices. We speak of a particle index if it
refers to a virtual single--particle state, i.~e.~a state which is not
occupied in a Slater determinant ground state whereas a hole index
refers to an occupied single--particle state. The particle--hole
part thus primarily describes excitations which, in a
single--particle picture, may be understood as lifting one fermion
from an occupied orbital to a virtual one. In a correlated system,
however, there are no fully occupied or fully virtual single--particle
states and thus the exact particle--hole part contributes also
for index pairs $(r, s)$ other than particle--hole index pairs.

The various approximation schemes may be classified by the order in
which the correlation is taken into account. Another criterion for
classification is which parts of the polarisation propagator are
included.

A simple first order approximation scheme based on $\Pi^{\ph}$ is the
so--called {\em Tamm--Dancoff Approximation (TDA)}
\cite{fetter71,ringSchuck}. It can be seen as the first order
approximation to the inverse matrix of the particle--hole
part.
In a quantum chemical language the TDA may be understood as treating
the excited states on the single--excitation Configuration Interaction
(SCI) level in comparison to an uncorrelated ground state.
The {\em Algebraic Diagrammatic Construction (ADC)} approach
\cite{schirmer82,trofimov97} represents a family of systematic higher
order approximation schemes to the particle--hole part.

Another group of approximations results from including the second part
of the polarisation propagator $\Pi^{\hp}_{rs,r's'}(\omega)$. This part is
called {\em hole--particle part} because its zeroth order
vanishes unless $r$ and $r'$ are hole indices and $s$ and $s'$ are
particle indices. In contrast to the particle--hole part its
poles appear at the negative excitation energies. The
fundamental first order approximation scheme which treats both parts
of the polarisation propagator on an equal footing is the so--called
{\em Random Phase Approximation (RPA)}
\cite{thouless61,rowe68:_eom,fetter71,ringSchuck}. Among the many
different ways of deriving this approximation we want to focus on the
role of the RPA as the first order approximation to the inverse matrix
of the polarisation propagator (\ref{pol_prop}). Naturally this matrix
now comprises the set of particle--hole as well as hole--particle
configurations and therefore has twice the dimension of the TDA
matrix. Among the higher order schemes that treat both parts of the
polarisation propagator we want to mention the {\em Second Order
Polarisation Propagator Approximation (SOPPA)}
\cite{nielsen80:_soppa,oddershede84,packer96:_soppa} and the {\em
Equation of Motion (EOM)} method \cite{rowe68:_eom,rose73:_eom}.

In the Configuration Interaction picture the RPA may be seen to
include ground state correlation in addition to the correlation of the
excited states already accounted for in the TDA
\cite{oddershede84,hansen78}. The RPA does so, however, in a
non--variational manner and thus it is not obvious that the RPA yields
improved results compared to the TDA. It has rather been observed that
in certain cases the RPA excitation energies are worse than those of
the TDA, also in systems where correlation of the ground state is of
special importance (for exemplary numerical comparisons see
e.~g.~\cite{koch95,bauernschmitt96}). In fact we will see later that
the influence of ground state correlation onto the RPA excitation
energies has to be regarded inconsistent with Rayleigh--Schr\"odinger
perturbation theory (see also Ref.~\cite{schirmer82}).  We will
further study a simple model system for which the RPA gives much
poorer results than the TDA. For this example a new approximation
method, called {\em First Order Static Excitation Potential (FOSEP)},
yields the exact solution while posing a matrix eigenvalue problem of
the same size as the RPA. Also a perturbation theoretical analysis
shows that FOSEP consistently includes ground state correlation.

The FOSEP method is a first order approximation scheme that
is derived from an {\em extended particle--hole Green's function}
\cite{brand_cederbaum_tbp} where additionally to the particle--hole
and the hole--particle parts other propagator terms are present:
The polarisation propagator has been augmented by combinations of
single--particle propagators that give rise to additional zeroth order
contributions in the particle--particle and the hole--hole index
space.  The introduced terms are chosen such that the extended Green's
function now satisfies a {\em Dyson equation} which in turn defines a
well--behaved {\em (particle--hole) self energy} in analogy to the
fundamental single--particle Green's function
\cite{fetter71,aTarantelli92,capuzzi96}. Earlier we have applied the
general procedure to the particle--particle propagator, another
well--known two--particle Green's function. We could show that the
self--energy of a suitably chosen extended two--particle Green's
function serves as an optical potential for elastic scattering of
two--particle projectiles
\cite{brand96}.  The particle--hole self energy, in turn, may be
understood as a sort of ``optical potential for particle--hole
excitations'', some general aspects of which are discussed in
Ref.~\cite{brand_cederbaum_tbp}.

This paper is organised as follows: After a brief review of the
relevant construction principles of the extended particle--hole
Green's functions we will define the FOSEP approximation and discuss
the structure of the corresponding matrix eigenvalue problem in
Sec.~\ref{sec_fosep}. In the next section the properties of the
FOSEP approximation for excitation energies and transition moments are
investigated on a formal level.  First the similarity to the RPA
equations is pointed out in order to compare the properties of the two
approximations (\ref{sec_rpa}). Both schemes share the fundamental
properties of size consistency (\ref{sec_size}) and the invariance
under unitary transformations of the occupied or virtual Hartree--Fock
orbitals (\ref{sec_unit}). The differences between the two
approximations will become apparent in Sec.~\ref{pert_exc} when we
will perform a perturbation theoretical analysis of the excitation
energies up to second order. This analysis shows that the FOSEP
approximation includes part of the ground state correlation in a
consistent way while the RPA proves inconsistent with perturbation
theory in this respect. It follows an analysis of the approximation
for transition moments which is found to be consistent in first
order. The last paragraph of Sec.~\ref{sec_prop} deals with the
equivalence of length and velocity form of the transition moments of
the dipole operator which are important for the so--called oscillator
strengths. In Sec.~\ref{sec_mod} the FOSEP method is compared with
the RPA and TDA in application to a simple exactly solvable model
system.


\section{The FOSEP Approximation} \label{sec_fosep}

The extended particle--hole Green's function is one species in a
family of two--particle propagators that fulfil Dyson's equation.  The
general theory has been developed in
Ref.~\cite{brand_cederbaum_tbp,brand96}. Here we will only outline the
main ideas that are relevant in the present context. The desired
physical information like excitation energies and transition moments
is contained in the poles and residues of the extended Green's
function.  Calculating these is equivalent to solving the eigenvalue
problem related to a generalised excitation energy operator $\cH$
which lives in an extended Hilbert space {\sf Y}.  Throughout this
paper we assume a discrete eigenvalue spectrum for the relevant
Hamiltonians since we are interested in finite basis set
approximations. The generalisation to continuous spectra, however,
poses no particular problem.

A basis of the extended Hilbert space {\sf Y} is given by the set of
states $\{\ket{Q_I}\}$ which are chosen ``orthonormal'' with respect
to an indefinite metric $\m$: \beq \bra{Q_I} \m \ket{Q_J} = \pm
\delta_{IJ} \eeq A model space is spanned within the full space by a
subset of basis states $\{\ky{rs}\}$. This model space is supposed to
allow the description of those excitations that are predominantly of
particle--hole type. Direct reference to `occupied' and `virtual'
single--particle states is avoided by allowing the indices $r$ and $s$
to range over the full set of single--particle indices each. The
states $\ky{rs}$ are chosen to include (ground state) correlation
being `correlated excited states' in the sense of
Refs.~\cite{mertins1,mertins2}. On the other hand they are constructed
manifestly `orthonormal', i.~e.~satisfying \beq \label{ortho_Y}
\by{rs} \m \ky{r's'} = \delta_{rr'} \delta_{ss'} \eeq exactly and in
each order of perturbation theory. It is a special property that all
states in the primary subset $\{\ky{rs}\}$ have positive norm. The
construction of states with the described properties presents the
crucial step in developing the theory of extended particle--hole
Green's functions.  Explicit expressions for the particular choice
used in this paper can be found in Ref.~\cite{brand_cederbaum_tbp}
together with a thorough discussion of the construction principles and
the remaining freedom of choice.

The basis $\{\ket{Q_I}\}$ defines a matrix representation
$\mat{\calH}$ of the generalised excitation energy operator $\cH$. The
subdivision of the basis into the basis of the model space
$\{\ky{rs}\}$ and the complementary part superposes a block structure
onto this matrix:
\begin{eqnarray} \label{block_H}
	{\mat{\calH}}& =& \left(\begin{array}{cc}
	\mat{\calH}_{aa}	&	\mat{\calH}_{ab}	\\
	\mat{\calH}_{ba}	&	\mat{\calH}_{bb}
	\end{array} \right) 
\end{eqnarray}
The index $a$ refers to the model space and $b$ to its
complement. The primary block $\mat{\calH}_{aa}$ of this matrix
is given by
\beqs \label{primary}
  \left[\mat{\cal H}_{aa}\right]_{rs,r's'} & = & \by{rs} \m \cH
    \ky{r's'}.
\eeqs
Explicit expressions for this matrix may be found in the appendix
and are derived in Ref.~\cite{brand_cederbaum_tbp} where also
a physical interpretation as a static particle--hole scattering
potential in the case of Coulomb interacting particles is given. In
the following we will investigate the first order approximation to this
matrix within the framework of many--body perturbation theory.

In order to apply perturbation theory the many--body Hamiltonian
$\hat{H}$ has to be split into two parts $\Ho$ and $\hat{H}_1$ as
usual. The choice of the one--particle operator
\beqs \label{def_Ho}
  \Ho  &=& \sum_i \varepsilon_i\, a_i^\dagger a_i
\eeqs
as characterised by the diagonalising single--particle basis
$\{\ket{\varphi_i}\}$ and single--particle energies
$\{\varepsilon_i\}$ defines the zeroth order of perturbation
theory. The residual interaction $\hat{H}_1 = \hat{v} + \hat{V}$
contains contributions of a one--particle operator $\hat{v}$ and a
two--body interaction $\hat{V}$:
\beqs \label{def_H1}
  \hat{v} &=& \sum_{i,j} v_{i j}\, a_i^\dagger a_j 
 \\  \hat{V} &=& 
     \frac{1}{2} \sum_{i,j,k,l} V_{i j k l}\, 
    a_i^\dagger a_j^\dagger a_l a_k
\eeqs

In particular we are interested in the M{\o}ller--Plesset partitioning
of the Hamiltonian where the zeroth order Hamiltonian $\Ho$ is defined
by the Hartree--Fock approximation. In the case of a nondegenerate
ground state the matrix elements $v_{ij}$ of the one--particle part of
the interaction are then given by
\beq \label{HFv}
  v_{ij}^{HF} = - \sum_k n_k V_{ik[jk]} .
\eeq
Here $V_{rs[r's']} = V_{rsr's'} -V_{rss'r'}$ denotes the
anti--symmetrised matrix element of the two--body interaction and
$n_r$ is the occupation number of the orbital $\ket{\varphi_r}$ in the
zeroth order ground state Slater determinant $\kphi$. 

The zeroth order of the primary block $\mat{\calH}_{aa}$ from
Eq.~(\ref{primary}) yields the matrix $\mat{\varepsilon}$ of zeroth
order excitation energies:
\beq \label{def_eps}
  \left[\mat{\cal H}_{aa}\right]^\zero_{rs,r's'} =
    \left[{\mat{\varepsilon}}\right]_{rs,r's'}  = (\varepsilon_r -
    \varepsilon_s)\, 
    \delta_{rr'} \delta_{ss'} 
\eeq
This reflects the fact that in our ansatz the first index in the pair
$rs$ has to be understood as marking the orbital (position) into
which a particle is created and the second index as marking the
orbital where a particle is destroyed (or a hole created).

The main motivation for developing the theory of the extended
particle--hole Green's function was that it fulfils a Dyson equation
and therefore possesses a particle--hole self energy $\mat{\Self} (\omega)$
\cite{brand_cederbaum_tbp,brand96}. This self energy may be seen to result
from a partitioning of the eigenvalue problem associated to the matrix
$\mat{\calH}$ with respect to the primary block $\mat{\calH}_{aa}$.
The energy independent (``static'') part
of the self energy $\mat{\Self} (\infty)$ is defined by the primary
block minus its zeroth order:
\beq
  \mat{\Self} (\infty) = \mat{\calH}_{aa} - {\mat{\varepsilon}}
\eeq
The significance of this part is to describe the the influence of
correlation to particle--hole excitations that remains in the high
energy limit, i.~e.~when the ``target particles have no time to
rearrange upon the influence of the particle--hole excitation''.  The
lowest order contributions to this matrix are of first order.  The
energy dependent (``dynamic'') part of the self energy $\mat{\Self}
(\omega) -\mat{\Self} (\infty)$ takes account of the remaining blocks
of the matrix $\mat{\calH}$ and starts in second
order.

The first order contributions to the (static) self energy are given by
\beqs \label{fo}
  \nonumber
  \left[\mat{\Self} (\infty)\right]^\first_{r s,r' s'} = \left[\mat{\cal
  H}_{aa}\right]^\first_{r s,r' s'} & = & v_{rr'} 
    \delta_{s s'} - v_{s's} \delta_{rr'} \\
  \nonumber
     & & + V_{rs'[sr']}\, (\nq_r n_s- n_r \nq_s)
         (\nq_{r'} n_{s'}- n_{r'} \nq_{s'}) \\
     & & + \delta_{s s'} \sum_k n_k V_{rk[r'k]}
         -  \delta_{r r'} \sum_k n_k V_{s'k[sk]}
\eeqs
for a general Hamiltonian
%
%
where we have introduced the notation $\nq_r = 1 - n_r$. 
Expression (\ref{fo}) 
may be readily derived from the general expression 
for the primary block $\mat{\calH}_{aa}$ which can be found in the
appendix.  In the M{\o}ller--Plesset case (\ref{HFv}), the first order
simplifies even further. In particular it is this approximation that
we will refer to as {\em First Order Static Excitation Potential
(FOSEP)}:
\beqs \label{def_S_fosep}
  \left[\mSsep\right]_{r s,r' s'} & = & V_{rs'[sr']}\, (\nq_r n_s- n_r \nq_s)
         (\nq_{r'} n_{s'}- n_{r'} \nq_{s'})
\eeqs
Now the following hermitian eigenvalue problem remains to be solved:
\beq \label{evp}
  \left( \mat{\varepsilon} + \mSsep \right) \underline{X} =
  \omega \underline{X}
\eeq
The (physical) eigenvalues $\omega$ provide approximations to the
excitation energies of the system and the corresponding eigenvectors
$\underline{X}$ may be used to calculate transition operator matrix
elements (transition moments). The transition moments corresponding to
the dipole operator define the so--called oscillator strengths which
are of great importance for photo--absorption and emission processes
\cite{oddershede84}. The FOSEP approximation for transition operator
matrix elements reads \cite{brand_cederbaum_tbp}
\beq\label{trans_mom}
  \bpsi \hat{T} \ket{\Psi_\mu}^\sep = \sum_{ij} T^{ij}
  X^\mu_{ij} ,
\eeq
where $T^{ij}$ are the matrix elements of the (one--particle)
transition operator $\hat{T}$ and $X^\mu_{ij}$ are the components
of the eigenvector associated to an excitation into the state
$\ket{\Psi_\mu}$.

Due to extensions included in the definition of the primary states
$\ky{rs}$, however, not all of the eigenvalues and eigenvectors of
Eq.~(\ref{evp}) correspond to ``physical'' excitations. More insight
may be gained by looking at the particular block structure of this
eigenvalue problem
which is revealed when splitting the set of index pairs according to
whether the indices relate to occupied (hole) or unoccupied (virtual
or particle) orbitals.  In zeroth order only the diagonal matrix
$\mat{\varepsilon}$ is present and the distinction between
``physical'' and ``unphysical'' excitations is obvious, since the
ground state is approximated by a Slater determinant. In this case
excitations of a single particle are possible only from an occupied
into a virtual orbital. Thus only the $ph$--$ph$ block is
``physical''. In the first order (FOSEP) approximation the secular
matrix has the following structure:
\beq \label{structure}
 \mat{\varepsilon} + \mSsep = 
   \begin{minipage}{13cm}

%
%

\unitlength2.5em
\begin{picture}(14,9)(2,0)
	\thicklines
	\put(4,0){\framebox(8,8){}}
	\thinlines
            \put(4,2){\line(1,0){8}}
            \put(4,4){\line(1,0){8}}
            \put(4,6){\line(1,0){8}}
            \put(6,0){\line(0,1){8}}
            \put(8,0){\line(0,1){8}}
            \put(10,0){\line(0,1){8}}
            \put(4,0){\makebox(2,2){0}}
            \put(4,2){\makebox(2,2){0}}
            \put(4,4){\makebox(2,2){0}}
            \put(4,6){\makebox(2,2){$\mat{\varepsilon}^{hh}$}}
            \put(6,2){\makebox(2,2){$\tilde{\mat{W}}^*$}}
            \put(6,0){\makebox(2,2){0}}
            \put(6,4){\makebox(2,2){$\mat{\varepsilon}^{ph} + \mat{V}$}}
            \put(6,6){\makebox(2,2){0}}
            \put(8,0){\makebox(2,2){0}}
            \put(8,2){\makebox(2,2){$-\tilde{\mat{\varepsilon}}^{ph} + \tilde{\mat{V}}^*$}}
            \put(8,4){\makebox(2,2){$\mat{W}$}}
            \put(8,6){\makebox(2,2){0}}
            \put(10,0){\makebox(2,2){$\mat{\varepsilon}^{pp}$}}
            \put(10,2){\makebox(2,2){0}}
            \put(10,4){\makebox(2,2){0}}
            \put(10,6){\makebox(2,2){0}}
            \put(3,0){\makebox(1,2){\footnotesize  $pp$}}
            \put(3,2){\makebox(1,2){\footnotesize  $hp$}}
            \put(3,4){\makebox(1,2){\footnotesize  $ph$}}
            \put(3,6){\makebox(1,2){\footnotesize  $hh$}}
            \put(4,8){\makebox(2,1){\footnotesize  $hh$}}
            \put(6,8){\makebox(2,1){\footnotesize  $ph$}}
            \put(8,8){\makebox(2,1){\footnotesize  $hp$}}
            \put(10,8){\makebox(2,1){\footnotesize  $pp$}}
\end{picture}

   \end{minipage}
\eeq
The asterisk ($^*$) denotes complex conjugation for the matrix elements
and the tilde ($\tilde{~}$) denotes a simultaneous transposition
of the two index pairs that specify a matrix element which simply
means a renumbering of the rows and columns of that matrix:
\[
  \left[\tilde{\mat{A}}\right]_{rs,r's'} = \left[{\mat{A}}\right]_{sr,s'r'}
\]

First of all we notice that the FOSEP self energy $\mSsep$ does not
contribute at all for pairs of orbitals that are both occupied ($hh$)
or both virtual ($pp$) and only the zeroth order matrix
$\mat{\varepsilon}$ remains. Thus the $hh$--$hh$ and $pp$--$pp$ blocks
decouple from the rest of the matrix and the eigenvalue problem
(\ref{evp}) for these blocks becomes trivial, simply yielding the
Hartree--Fock orbital energy differences.
These blocks are obviously not correlated in the first order treatment.
We want to mention that this decoupling of the $hh$--$hh$ and
$pp$--$pp$ blocks is special to the M{\o}ller--Plesset partitioning of
the Hamiltonian and to the particular choice of the extended states
$\ky{rs}$ considered in this paper. It does not appear for other
choices discussed in Ref.~\cite{brand_cederbaum_tbp}. The
decoupling leads to a considerable reduction of numerical effort and
therefore justifies the present choice. Many of the properties
discussed in the present paper, however, generalise also to first
order approximations based upon other choices for the primary
extended states $\ky{rs}$.

The $ph$ block of the $\mat{\varepsilon}$ matrix
$\mat{\varepsilon}^{ph}$ contains those energies that
relate to a simple particle--hole excitation in a zeroth order
picture. The contribution of 
\beq \label{def_V_block}
  \left[\mat{V}\right]_{ph,p'h'} = V_{ph'[hp']}
  \qquad\qquad\mbox{($h$, $h'$ occupied, $p$, $p'$ virtual orbitals)}
\eeq
in the $ph$--$ph$ block describes the interaction of the uncorrelated
ground state Slater determinant with a singly excited configuration.
In fact, diagonalising the $ph$--$ph$ block $\mat{\varepsilon}^{ph} +
\mat{V}$ on its own results in the well--known {\em Tamm--Dancoff
Approximation (TDA)} \cite{fetter71}. The coupling $\mat{W}$ to the
$hp$--$hp$ block can be understood as taking into account ground state
correlation as will be explained later with the help of perturbation
theoretical arguments. A similar coupling also appears in the {\em
Random Phase Approximation (RPA)} \cite{fetter71}. The relation of our
approach to the RPA will be discussed in detail below.

The $hp$--$hp$ block by itself does not seem very physical in
character at all. Its zeroth order excitation energies are negative
and result from creating a hole in a virtual orbital and a particle in
an occupied orbital. Nevertheless, the $hp$--$hp$ block couples
through the matrix
\beq \label{def_W_block}
  \left[\mat{W}\right]_{ph,h'p'} = - V_{pp'[hh']}
  \qquad\qquad\mbox{($h$, $h'$ occupied, $p$, $p'$ virtual orbitals)}
\eeq
with the physical $ph$--$ph$ block and thus introduces a correction to
the Tamm--Dancoff excitation energies.  Due to the decoupling of the
$pp$ and $hh$ blocks we are left with an eigenvalue problem comprising
the blocks of the FOSEP matrix with $ph$ and $hp$ indices:
\beqs \label{Mevp}
  \mMsep\, \underline{x} &=&
  \omega\, \underline{x} \\ \label{def_Msep}
  \mMsep &=& \mat{\varepsilon} + {\mSsep} \bigg{|}_{
  \mbox{\scriptsize $ph$ and $hp$ blocks}} = \left( 
  \begin{array}{cc} 
     \mat{\varepsilon}^{ph} + \mat{V} & \mat{W} \\ 
     \tilde{\mat{W}}^* & - \tilde{\mat{\varepsilon}}^{ph} + \tilde{\mat{V}}^*
  \end{array} \right)
\eeqs
This is the eigenvalue problem that has to be solved in the FOSEP
approximation scheme. In contrast to the RPA it is a hermitian
eigenvalue problem always yielding real eigenvalues. In fact, in most
cases the matrix $\mMsep$ is real symmetric.

As long as the interaction remains weak enough, there is a clear
distinction between ``physical'' eigenvalues of this matrix and
``unphysical'' ones by the sign of these energies. Even with stronger
interaction the distinction may still be valid. This is understood
easily when considering a model system where the $ph$--$ph$ block has
the dimension one and all matrix elements are real. The matrix $\mMsep$
then is two by two and its eigenvalues are given by
\[
  \pm \sqrt{{\varepsilon^{ph}}^2 + W^2} + V.
\]
Thus we get one positive eigenvalue and a negative one, provided
that
\[
  {V^2} < {{\varepsilon^{ph}}^2 + W^2} .
\]
Note that $V$ is the contribution of the correlation introduced in the
TDA and this condition now states that it has to be small enough
compared to the zeroth order excitation energy augmented by the
additional interaction term $W$.  When this condition is violated, or
in general the numbers of positive and negative eigenvalues in a given
symmetry are not the same, the eigenvectors may be necessary to
distinguish between physical and unphysical contributions. Still the
``physical'' approximation may usually be defined by the
upper half of the eigenvalues.

In the remaining sections of this paper we will discuss the FOSEP
approximation as defined above. It presents the natural first step in
approximating the particle--hole self energy $\mat{\Self} (\omega)$
and thus the matrix $\mat{\calH}$. At this place we want to mention
that other approximations, for example, result by augmenting the
primary set of states $\ky{rs}$. In particular one can obtain the RPA
straightforwardly via the formalism of extended Green's
functions. This is achieved by additionally including a subset of the
basis $\{\ket{Q_I}\}$ consisting of states with negative norm that are
degenerate (in zeroth order) to the hole--particle fraction of the set
$\{\ky{rs}\}$ \cite{brand_cederbaum_tbp}. This augmented set of states
defines an extension of the primary block $\mat{\calH}_{aa}$ in the
matrix $\mat{\calH}$. In first order this extended matrix can be
decoupled with the help of a unitary transformation into the RPA
eigenvalue problem and additional unphysical blocks. Thus the RPA is
included in the general theory as a specific approximation.  Note,
however, that FOSEP presents the {\em canonical} first order
approximation in our ansatz since it is based upon the primary set of
states $\{\ky{rs}\}$ which defines the Green's function and self
energy matrices.


\section{Properties of the FOSEP Approximation} \label{sec_prop}

In this chapter we will discuss some general properties of the FOSEP
approximation. In order to elucidate the relation to the well known
first order approximation schemes RPA and TDA we start with briefly
reformulating the RPA in our notation. We go on with considering two
fundamental invariances FOSEP shares with RPA and TDA namely size
consistency and invariance with respect to unitary transformations of
the single--particle basis. In order to show the differences between
the three schemes we will carry out a perturbation theoretical
analysis for the excitation energies as well as for the
transition moments. Finally, the equivalence of length and velocity
form of the dipole operator transition moments is discussed.

\subsection{Relation to the RPA} \label{sec_rpa}

The {\em Random Phase Approximation (RPA)} \cite{fetter71,ringSchuck} for the
calculation of excitation energies and transition operator matrix
elements in finite Fermi systems may be derived and understood in
many different ways. 
Traditionally the RPA is derived
by the infinite summation of a certain type of diagrams in the
Feynman--Dyson perturbation series of the polarisation propagator
\cite{thouless61}. Equivalently it can be understood
as a first order approximation to the integral kernel of the
Bethe--Salpeter equation \cite{czyz61} or as a specific
first order approximation in the equation of motion of the
polarisation propagator \cite{rowe68:_eom,oddershede84}. Now we are
going to present the RPA equations in a form suitable for comparison
with FOSEP.

Based upon a Hartree--Fock zeroth order the RPA is defined by the
following eigenvalue problem \cite{schirmer96}:
\beq
  \left( \mat{\varepsilon}\, \mat{m} + \mat{R} \right) \underline{x} = 
  \omega\, \mat{m}\, \underline{x}
\eeq
The matrix $\mat{\varepsilon}$ is defined like above in
Eq.~(\ref{def_eps}).  The RPA kernel $\mat{R}$ consists of the matrix
elements $R_{r s,r' s'} =  V_{rs'[sr']}$ and the metrical
matrix $\mat{m}$ by
\beq
  \left[\mat{m}\right]_{rs,r's'} = \delta_{r r'} \delta_{s s'}\,
  (\nq_r n_s- n_r \nq_s) .
\eeq
Note that in block matrix notation $\mat{m}$ can be written as
\beq
  \mat{m} = \left( \begin{array}{cc}
               \mat{1} & \mat{0} \\
               \mat{0} & -\mat{1}
            \end{array} \right) .
\eeq
All these matrices are indexed by pairs of single--particle indices
that are required to be either particle--hole or hole--particle index
pairs. Thus the RPA eigenvalue problem has the same size as the FOSEP
one (\ref{Mevp}). The essential difference is the appearance of the
indefinite metric $\mat{m}$ in the RPA case which renders the RPA
problem a non--hermitian eigenvalue problem. As a consequence the RPA may
become instable and lead to complex eigenvalues \cite{oddershede84}.  The
RPA kernel $\mat{R}$ is related to the FOSEP self energy $\mSsep$ in
the following way:
\beq
  \mSsep = \mat{m}\,\mat{R}\,\mat{m}
\eeq
Introducing the matrix
\beq \label{def_Mrpa}
  \mMrpa = \mat{\varepsilon}\,\mat{m} + {\mSsep} \bigg{|}_{
  \mbox{\scriptsize $ph$ and $hp$ blocks}} = \left( 
  \begin{array}{cc} 
     \mat{\varepsilon}^{ph} + \mat{V} & \mat{W} \\ 
     \tilde{\mat{W}}^* & \tilde{\mat{\varepsilon}}^{ph} + \tilde{\mat{V}}^*
  \end{array} \right)
\eeq
with the same nomenclature as in (\ref{def_V_block}) and (\ref{def_W_block}),
the RPA eigenvalue problem may be rewritten to
\beq \label{rpa_evp}
   \mMrpa \, \underline{x}' =
  \omega\, \mat{m} \,\underline{x}'
\eeq
with $\underline{x}' = \mat{m} \,\underline{x}$. Comparing the RPA
(\ref{rpa_evp}, \ref{def_Mrpa}) with the FOSEP eigenvalue problem
(\ref{Mevp}, \ref{def_Msep}) we see that both have the same size and
start from the same input data while the difference lies in some minus
signs. Before analysing the differences further we want to discuss two
fundamental properties that are shared by both schemes.

\subsection{Size Consistency of FOSEP} \label{sec_size}

The question of size consistency of a many--body method is the
question of whether the resulting approximations for physical
quantities scale correctly with the size of the system
\cite{march67,mcweeny89}. The general question is difficult to
answer and usually one has to resort to simple models or numerical
calculations. Nevertheless, this concept becomes very important for
applications to large or extended systems.  In the context of finite
systems, especially molecules, the so--called separate fragment model
provides a useful test of correct scaling behaviour. We consider a
many--body system consisting of two or more separate (noninteracting)
subsystems (fragments). Size consistency of excitation energies and
transition moments then means that an excitation that is local to one
of the subsystems is approximated with the same result regardless
whether the approximation scheme is applied to the full system or only
to the fragment. A sufficient but not necessary condition for this
property is that the secular equations of the approximation scheme
give rise to independent sets of equations corresponding to local
excitations on the individual subsystems. This `a priori' decoupling 
of independent, local equations is known as the separability property
\cite{schirmer96:size_consistency}. 

The separability of FOSEP and RPA is proven by the following
arguments: In the model of separating fragments the Hamiltonian of the
full system is given by the sum of the Hamiltonians of the
subsystems. This implies that the (Hartree--Fock) single--particle
states $\varphi_r$ may be chosen local to either fragment and that the
matrix elements of the two--body interaction $V_{ijkl}$ vanish unless all
indices $ijkl$ relate to states belonging to the same subsystem. 
From the definition of the FOSEP matrix (\ref{def_S_fosep},
\ref{def_Msep}) and the RPA matrix (\ref{def_Mrpa}) it therefore
becomes clear that both methods have the separability property and
thus can be regarded size consistent. The same arguments apply to the
TDA to which the FOSEP and RPA reduce in the case of vanishing
coupling $\mat{W}$ as explained in Sec.~\ref{sec_fosep}.  For
nonlocal excitations the excitation energy is simply given by the
difference of the single--particle (Hartree--Fock) energies for all
three schemes. This means that the level of approximation is that of
Koopman's theorem which provides a consistent first order description.
We want to mention that the separability property is by no means a
matter of course for more accurate many--body methods like, for
example, the general CI method \cite{szabo82:_moder_quant_chemis}.

\subsection{Unitary Transformations of Single--Particle Orbitals}
\label{sec_unit} 

In order to separate the influence of finite truncations of the
underlying single--particle basis from the systematic deficiencies of
a given approximation scheme it is important that the approximation is
invariant with respect to rotations of the single--particle basis.  A
global invariance is self--evident only for `exact' methods like full
CI. Systematic truncations of the CI matrix employing single, double,
or triple excitations on a given reference configuration are at least
invariant with respect to transformations of the single particle basis
that do not mix occupied and virtual orbitals. Such an invariance
usually does not apply for perturbative propagator methods. The FOSEP
method as well as the RPA and the TDA, however, share this invariance
with the CI whereas higher order methods usually do not. For the SOPPA
method the influences of rotations of the orbital set have been
investigated numerically \cite{oddershede83}.

A physical motivation for altering the single--particle functions may
be drawn from the fact that the Hartree--Fock virtual orbitals
describe additional test particles in a mean field and therefore
constitute rather diffuse functions while the main effects of
correlation show up at short range due to ineffective screening. Thus
one can hope to achieve a better description of the influence of
correlation with more localised virtual orbitals than the
Hartree--Fock ones.

Within a perturbation theoretical approach a unitary transformation
within the set of virtual single--particle states can be realised by
adding a (hermitian) single--particle potential to $\Ho$ of
Eq.~(\ref{def_Ho}) that takes effect only on the virtual orbitals and
subtracting it again from $\hat{H}_1$ of Eq.~(\ref{def_H1}). The new
single--particle basis is then defined as the diagonalising basis for
the new zeroth order Hamiltonian. Obviously the new basis is connected
to the original one by a unitary transformation that leaves invariant
the occupied single--particle states and also does not affect the
Slater determinant $\kphi$ preserving the distinction between occupied
and virtual single--particle states.  From the definitions of the
matrices (\ref{def_eps}, \ref{def_S_fosep}) it can be seen that such a
transformation of the single--particle basis also results in a unitary
transformation of the secular matrix in (\ref{evp}) which preserves
its block structure (\ref{structure}). Therefore also the FOSEP
eigenvalue problem (\ref{Mevp}, \ref{def_Msep}) transforms without
changing its eigenvalues. This argumentation can be transferred
analogously to the closely related RPA and TDA. Summarising, we have
seen that the FOSEP approximation as well as the RPA and TDA are
invariant under unitary transformations within the set of virtual
single--particle states.  It is easily seen that this property
generalises to unitary transformations of the orbital basis that do
not mix occupied and virtual orbitals.

In a much more general sense, however, the matrix $\mat{\cal H}_{aa}$
of Eq.~(\ref{primary}) which forms the primary block of the matrix
representation $\mat{\cal H}$ of the excitation energy operator
$\hat{\cal H}$ is invariant under (unrestricted) unitary
transformations of the single particle space. Since the only unknown
quantity in $\mat{\cal H}_{aa}$ (see also Eq.~(\ref{H_aa_app}) in the
appendix) is the exact ground state of the system, the invariance
properties of an approximation to $\mat{\cal H}_{aa}$ follow the
chosen approximation for the ground state. In other words: The
eigenvalues of the matrix of Eq.~(\ref{H_aa_app}) depend {\em only} on
the chosen approximation for the ground state and {\em not} on the
particular choice of the single--particle basis. In the special case
of a system of particles which interact only with a one--particle
potential, the primary block $\mat{\cal H}_{aa}$ is even independent
of the ground state. In this case the first order already provides the
exact solution for the excitation energies and is invariant with
respect to any unitary transformations of the single--particle
basis. This property is explained in more detail in 
Ref.~\cite{brand_cederbaum_tbp}.

\subsection{Perturbation Theoretical Analysis of the Excitation
Energies} \label{pert_exc}

In order to analyse the differences between the FOSEP approximation,
the RPA, and the TDA we will now perform a perturbation theoretical
analysis of the excitation energies up to second order and compare
with straightforward Rayleigh--Schr\"odinger perturbation theory
following Ref.~\cite{schirmer82}.  Usually the
Rayleigh--Schr\"odinger series itself is not a reliable method for
calculating energies of excited states but it is very helpful for
analysing and comparing different approximation schemes.

We assume that the Rayleigh--Schr\"odinger series starting from the
singly excited Slater determinant
\beq
  \ket{\Phi_{\alpha \beta}} = a_\alpha^\dagger a_\beta \kphi
\eeq
converges towards the excited state $\ket{\Psi_{\alpha \beta}}$. Note
that here $\alpha$ has to be a particle index and $\beta$ a hole index.

An expression for the second order excitation energy can be gained by
subtracting the expressions for the ground state energy $E_0$ from the
excited state energy $E_{\alpha \beta}$. Up to second order the
ground state energy is given by the familiar expression
\beq \label{second_E_0}
  E_0{(2)} = E_0{(1)} + U_0^{(2p-2h)} ,
\eeq
where $E_0{(1)} = \bphi \hat{H} \kphi$ is the first order ground state
energy. The term
\beq \label{def_U_0}
  U_0^{(2p-2h)} = - \sum_{i < j \atop k < l} \frac{\left| V_{ij[kl]}
  \right|^2}{\varepsilon_i + \varepsilon_j -\varepsilon_k
  -\varepsilon_l} \overline{n}_i \overline{n}_j n_k n_l 
\eeq
denotes the second order contribution to the ground state
correlation. The given approximation is known as the M{\o}ller--Plesset
(MP)--2 approximation and extensively used in quantum chemistry. In the 
Configuration Interaction language the second order term $U_0^{(2p-2h)}$ may
be interpreted to present interactions of the ground state Slater determinant
$\kphi$ with two particle--two hole configurations.

The second order energy of the excited state is also evaluated
straightforwardly and can be found in Ref.~\cite{schirmer82}. The
excitation energy up to second order then reads
\beqs \label{exc_energy_rspt2}
  \nonumber
  \Delta E_{\alpha \beta}{(2)}& =& E_{\alpha \beta}{(2)} - E_0{(2)}\\
  & =&  \Delta E_{\alpha \beta}{(1)} 
  + U_{\alpha \beta}^{(p-h)} + U_{\alpha \beta}^{(2p-2h)} + R_{\alpha \beta} 
\eeqs
where $\Delta E_{\alpha \beta}{(1)} = \varepsilon_\alpha -
\varepsilon_\beta - V_{\alpha \beta[\alpha \beta]}$ is the first order
excitation energy. The terms $ U_{\alpha \beta}^{(p-h)}$ and
$U_{\alpha \beta}^{(2p-2h)}$ denote second order contributions to the
excited state's energy arising from the interaction of the
configuration $\ket{\Phi_{\alpha \beta}}$ with (other) $p-h$ and
$2p-2h$ configurations, respectively, and can be found in
Ref.~\cite{schirmer82}.
The part $R_{\alpha \beta}$ is the remainder of a partial cancellation
of the second order ground state correlation term $U_0^{(2p-2h)}$ of
Eq.~(\ref{def_U_0}) with a contribution to the correlation of the
excited state. It can be written as the sum of three terms
\beq
  R_{\alpha \beta} = R^1_{\alpha \beta} + R^2_{\alpha \beta} + R^3_{\alpha \beta}
\eeq
where 
\beqs \label{three_parts_R}
  \nonumber
  R^1_{\alpha \beta} &=& \sum_{j, k, l \atop {k < l \atop k, l \neq \beta}}
  \frac{\left| V_{\alpha j[kl]} \right|^2}{\varepsilon_\alpha + \varepsilon_j
  -\varepsilon_k -\varepsilon_l} \overline{n}_j n_k n_l \\
   R^2_{\alpha \beta} &=& \sum_{i, j, l \atop {i < j \atop i, j \neq \alpha}}
  \frac{\left| V_{ij[\beta l]} 
  \right|^2}{\varepsilon_i + \varepsilon_j -\varepsilon_\beta
  -\varepsilon_l} \overline{n}_i \overline{n}_j n_l \\
  \nonumber
   R^3_{\alpha \beta} &=& \sum_{j, l} \frac{\left| V_{\alpha j[\beta l]}
  \right|^2}{\varepsilon_\alpha + \varepsilon_j -\varepsilon_\beta
  -\varepsilon_l} \overline{n}_j n_l
\eeqs
These contributions are left over from $U_0^{(2p-2h)}$ corresponding to
the special cases where $i = \alpha$ or $k = \beta$ in the sum of
Eq.~(\ref{def_U_0}). The rest of $U_0^{(2p-2h)}$ is cancelled by
contributions from the excited state.

We are now in the position to compare with the second order excitation
energies from the TDA, the RPA, and the FOSEP scheme. The
approximations for the excitation energies in these schemes are found
by solving the eigenvalue problem related to the corresponding
matrix. Basic matrix perturbation expansion leads to the second order
approximation for the eigenvalue.
We find the following:
\beqs
  &{\Delta E^\tda_{\alpha \beta}}{(2)} & = \Delta E_{\alpha \beta}{(1)}+
  U_{\alpha \beta}^{(p-h)} \\
  &{\Delta E^\rpa_{\alpha \beta}}{(2)} & = \Delta E_{\alpha \beta}{(1)}+
  U_{\alpha \beta}^{(p-h)} - R^3_{\alpha \beta} \\
  &{\Delta E^\sep_{\alpha \beta}}{(2)} & = \Delta E_{\alpha \beta}{(1)}+
  U_{\alpha \beta}^{(p-h)} + R^3_{\alpha \beta}
\eeqs
All three approximation schemes are consistent in first order with the
Rayleigh--Schr\"odinger expression (\ref{exc_energy_rspt2}). Therefore
they are correctly referred to as first order schemes.  The zeroth and
first order contribution $\Delta E_{\alpha \beta}{(1)}$ originate in
the diagonal matrix elements of the $ph$--$ph$ block (TDA block) of
the matrices $\mMsep$ and $\mMrpa$ of Eqs.~(\ref{Mevp},
\ref{def_Mrpa}) whereas the second order terms come in by the
diagonalisation procedure. Neither of the three schemes reproduce the
second order expression (\ref{exc_energy_rspt2}) completely. This is
only achieved by more accurate and more costly schemes like ADC(2) or
SOPPA.

The term $U_{\alpha \beta}^{(p-h)}$ describes part of the second order
correlation of the excited state as can be seen from
Eqs.~(\ref{exc_energy_rspt2}) and (\ref{second_E_0},
\ref{def_U_0}). It is the only second order contribution to the TDA
excitation energy and originates in the off--diagonal part of the TDA
matrix which also constitutes the $ph$--$ph$ block of the
FOSEP matrix $\mMsep$.
In the FOSEP and the RPA expressions the additional term $R^3_{\alpha
\beta}$ is present which has already been identified as part of the
ground state correlation. Comparing with the second order perturbation
expansion (\ref{exc_energy_rspt2}), however, we notice that it appears
with the wrong sign in the RPA excitation energy in contrast to FOSEP
where the sign is consistent with the Rayleigh--Schr\"odinger
expansion. This term comes in through the coupling of the $ph$ with
the $hp$ block in the FOSEP matrix $\mMsep$ of Eq.~(\ref{Mevp}) and
the RPA matrix $\mMrpa$ of Eq.~(\ref{def_Mrpa}), respectively. We
recall that the only difference between the FOSEP and the RPA
equations are minus signs in the $hp$--$hp$ blocks of the secular
matrices and the RPA metric. These carry through to the second order
expressions and prove inconsistent with Rayleigh--Schr\"odinger
perturbation theory in the RPA case.

It is interesting to note that each of the three parts of $R_{\alpha
\beta}$ of Eq.~(\ref{three_parts_R}) is positive. Thus the RPA always
lowers the TDA excitation energies in second order whereas the FOSEP
approximation increases the energies in accordance with the positive
sign of the full correction $R_{\alpha \beta}$. Taking into account
that the TDA approximates the ground state by the variational
Hartree--Fock method, it is sensible to expect an increase in the
excitation energies when ground state correlation is additionally
taken into account. Therefore we may conclude that the FOSEP scheme
includes ground state correlation in a consistent way in contrast to
the RPA which does not.

Concerning the term $U_{\alpha \beta}^{(2p-2h)}$ in the full second
order expression (\ref{exc_energy_rspt2}) which is missing in all
three schemes, we remark that it does not carry a definite sign (as
can be seen from Eq.~(68c) of Ref.~\cite{schirmer82}). It may,
however, lower the energy of the excited state and thus can possibly
overcompensate for the influence of the ground state correlation. In
this way the RPA result may be supported by accidental numerical
compensation.

\subsection{Perturbation Theoretical Analysis of the Transition Moments}

We now will show that transition operator matrix elements between the
ground state and an excited state can, in contrast to TDA, be
approximated by FOSEP consistently in first order.  We consider the
transition moment
\beq
  {\cal T}_{\alpha \beta} = \bpsi \hat{T} \ket{\Psi_{\alpha \beta}} 
\eeq
of the (one--particle) transition operator $\hat{T}$ for the
particle--hole excited state $\ket{\Psi_{\alpha \beta}}$ that was introduced in
the last paragraph. Up to first order the perturbation expansion yields
\beqs
  {\cal T}_{\alpha \beta}{(1)}& =& \bphi \hat{T} \ket{\Phi_{\alpha \beta}} +
  \bphi \hat{T} \ket{\Psi^{(1)}_{\alpha \beta}}  +
  \bra{\Psi^{(1)}_{o}} \hat{T} \ket{\Phi_{\alpha \beta}} .
\eeqs
Explicit expressions for these terms derive from straightforward
Rayleigh--Schr\"odinger perturbation theory and may be found in
Ref.~\cite{schirmer82}. 

In an analogous fashion to the preceding paragraph the above
expression may be compared to the result of matrix perturbation
analysis of the TDA, RPA, and FOSEP approximations for the transition
moments. It is easily seen that the FOSEP eigenvalue problem
(\ref{evp}, \ref{Mevp}) together with the approximation for the
transition moments of Eq.~(\ref{trans_mom}) leads to a consistent
first order approximation of the transition moments. For the TDA, the
RPA, and the first and second order ADC scheme the perturbation
analysis has been carried out in Ref.~\cite{schirmer82}. It has been
found that the TDA expression is incomplete in first order because the
term $\bra{\Psi^{(1)}_{o}} \hat{T} \ket{\Phi_{\alpha \beta}}$ is
missing. This term is a consequence of first order ground state
correlation which is neglected in the TDA. It is described correctly
in the RPA which yields consistent transition moments through first
order. This may be seen to justify the common statement that the RPA,
in contrast to the TDA, considers ground state correlation. We want to
mention that the transition moments are also described consistently
through first order in the ADC(1), an approximation that is equivalent
to the TDA with respect to the excitation energies but differs for the
transition moments \cite{schirmer82}.

\subsection{The Equivalence of Length and Velocity Form of the Dipole
Operator Transition Moments}

Now we will focus on the transition moments of the dipole operator as
a particular choice for the transition operator. These transition
moments are related to the so-called oscillator strengths and present
important parameters characterising the interaction of the many--body
system with radiation. 
There is a hierarchy of equivalent representations of the exact dipole
operator transition moments, starting with the so--called length and
velocity form. Without loss of generality we consider only the
$z$--component $\hat{Z}$ of the dipole operator. The identity
\beq \label{eqv_l_v}
  (E_\mu - E_0) \bra{\Psi_\mu}\hat{Z}\kpsi =
   - i \bra{\Psi_\mu} \hat{P}_z\kpsi 
\eeq
expresses the equivalence between the length and the velocity form of
the dipole operator transition moments defined by the left and the
right hand side of this equation, respectively. The $z$--component of
the momentum operator $\hat{P}_z$ is related to the dipole operator by
$ \left[\hat{H}\, , \, \hat{Z}\right] = - i
\hat{P}_z$ provided that the Hamiltonian contains only local
potentials like e.~g.\ for Coulomb interacting electrons in atoms or
molecules.  Eq.~(\ref{eqv_l_v}) then follows from the identity
$(E_\mu - E_0)
\bra{\Psi_\mu}\hat{Z}\kpsi =\bra{\Psi_\mu}\left[\hat{H}\, , \,
\hat{Z}\right]\kpsi$.

It is a very special property of the RPA to preserve this equivalence
exactly, provided the underlying Hartree--Fock single--particle basis
is complete \cite{harris69}. In fact the RPA eigenvalue problem may be
derived setting out from a CI representation of the ground and excited
state wavefunction and requiring certain `hypervirial
relations' which present a slight generalisation of the above mentioned
equivalences \cite{hansen78}. Thus the exact fulfilment of these
hypervirial relations may be regarded unique to the RPA while other
approximation schemes show, at most, a perturbative equivalence.
In order to make the perturbation expansion transparent, it is useful
to introduce the function $\Delta(\lambda)$ as the difference between the
right and the left hand side of Eq.\ (\ref{eqv_l_v}). It is a function
of the usual interaction strength parameter $\lambda$ in $\hat{H}_\lambda =
\hat{H}_0 + \lambda \hat{H}_1$. Owing to its non--local character,
Hartree--Fock does not preserve the length--velocity
equivalence. Hence, choosing $\hat{H}_0$ to be the Hartree--Fock
operator, $\Delta(\lambda)$ does not vanish for any $\lambda$ except for
$\lambda =1$.
The expansion of $\Delta(\lambda)$ in powers of $\lambda$ does not vanish
term by term; each order will in general give a nonvanishing
contribution. The TDA approximates only the zeroth order term of
$\Delta(\lambda)$ correctly, giving an error in first order [i.\ e.\
$\cal{O}(\lambda)$]. The FOSEP approximates $\Delta(\lambda)$ through first
order giving a second order error. So does the RPA which additionally
has the unique property of reproducing the zero $\Delta(\lambda =1) =
0$. For completeness we mention that the first order ADC
\cite{trofimov95} approximates $\Delta(\lambda)$ through first order
while the SOPPA \cite{jorgensen83} as well as the second order ADC
\cite{trofimov95} are consistent through second order.

The result of equivalence
through first order is another clue for the consistency of the FOSEP
approximation. The 
error between length and velocity form
provides a genuine test for the validity of the
approximation. Depending on the particular aims of an approximate
calculation this may seem more favourable than having an `a priori'
equivalence like in the RPA which hides the actual error with respect
to the exact value of the transition moments.


\section{Application to a Simple Model} \label{sec_mod}

In this section we compare the FOSEP approximation with the RPA, the
TDA, and the exact solution for the excitation energies of a very
simple model system. This model is commonly referred to as the Hubbard
model for the Hydrogen Molecule H$_2$
\cite{ashcroft76:_solid_state_physic}. In the model, all excitation
energies can be calculated by analytic expressions in dependence of
two parameters which mimic the effects of Coulomb interaction.

\subsection{Definition of the Model and Exact Solution}

The Hubbard model for the diatomic Hydrogen molecule represents each
atomic site ({\bf R} and {\bf R'}) by a single orbital electronic
level denoted by $\ket{\mbox{\bf R}}$ and $\ket{\mbox{\bf R'}}$,
respectively. Each level can be occupied by up to two electrons with
opposite spin. The single--particle part of the Hamiltonian $\hat{h}$
contains a diagonal term
\beq
  \bra{\mbox{\bf R}} \hat{h} \ket{\mbox{\bf R}}=
  \bra{\mbox{\bf R'}} \hat{h} \ket{\mbox{\bf R'}}= {\cal E}
\eeq
which yields an energy ${\cal E}$ for each electron. The off--diagonal
term describes attraction by the neighbouring nucleus and represents
an amplitude for `tunnelling' or `hopping' of an electron from one site
to another:
\beq
  \bra{\mbox{\bf R}} \hat{h} \ket{\mbox{\bf R'}}=
  \bra{\mbox{\bf R'}} \hat{h} \ket{\mbox{\bf R}}= -t \quad (< 0)
\eeq
Additionally a two--particle interaction term is present which yields
a positive energy $U$ whenever one level is occupied by two
electrons. This term represents the intra--atomic Coulomb repulsion
between two localised electrons. Note that all interactions are
independent of electron spin. 

We consider the neutral H$_2$ molecule, i.~e.~an occupation with two
electrons.  The solution of the Hartree--Fock equations yields the
(molecular) orbital functions
\beq \label{mol_orb_g}
  \ket{g} = \frac{1}{\sqrt{2}}(\ket{\mbox{\bf R}} + \ket{\mbox{\bf
  R'}}) \qquad \qquad
  \ket{u} = \frac{1}{\sqrt{2}}(\ket{\mbox{\bf R}} - \ket{\mbox{\bf
  R'}}).
\eeq
while the corresponding Hartree--Fock single--particle energies read
$\varepsilon_{g/u} = {\cal E} \mp t + \frac{1}{2} U$. In the
Hartree--Fock ground state Slater determinant $\kphi = \ket{g\uparrow
g\downarrow}$ the orbital $\ket{g}$ is doubly occupied. In order to
make transparent the transition between the Hartree--Fock approximation
and the correlated problem we introduce the additional (perturbation)
parameter $\lambda \in [0,1]$ by using the Hamiltonian
\beq
  \hat{H}_\lambda = \Ho + \lambda \hat{H}_1 
\eeq
where the Fock operator $\Ho$ and the interaction Hamiltonian
$\hat{H}_1$ are defined as in Sec.~\ref{sec_fosep}. The matrix
elements of the two--body interaction $V_{ijkl}$ that define
$\hat{H}_1$ are given by the transformation (\ref{mol_orb_g}) into the
atomic orbital picture. 

There are six independent solutions for the two--electron eigenstates
of the system. According to the possible combinations of the
electrons' spins three states of singlet symmetry
$\ket{\Psi_{0/1}}$ and $\ket{S_u}$ and a degenerate triplet
$\ket{T_{-1/0/1}}$ are found. The triplet and the singlet $\ket{S_u}$
states are uncorrelated and have one electron in a $u$ and one in a $g$
orbital. The singlet state $\ket{\Psi_{1}}$ corresponds to a $2p-2h$
excitation. It will no longer be considered because it lies outside of
the range of RPA, TDA, and FOSEP. 
The wavefunction of the singlet ground
state is given by 
\beqs
  \ket{\Psi_0}_\lambda & =& \alpha_\lambda \ket{g\uparrow
  g\downarrow} + \beta_\lambda \ket{u\uparrow   u\downarrow}
\eeqs
where
\beq \label{aph_und_beth} \begin{array}{lcl}
  \alpha_\lambda &=& { \displaystyle  {{4\,t + {\sqrt{16\,{t^2} + {{\lambda }^2}\,{U^2}}}}\over 
    {{\sqrt{{{{\lambda }^2}\,{U^2}} + {{{{\left( 4\,t + 
                  {\sqrt{16\,{t^2} + {{\lambda }^2}\,{U^2}}} \right) }^2}}
              }}}}}} \\
  \beta_\lambda^2 & = & 1 - \alpha_\lambda^2 .
\end{array} \eeq
The dependence on the perturbation parameter $\lambda$ indicates the
influence of correlation which only appears 
between the singlet Slater determinants 
of $g$ symmetry
$\ket{g\uparrow g\downarrow}$ and $\ket{u\uparrow u\downarrow}$.
In order to simplify the notation, we will drop the subscript
$\lambda$ in the following. Note that the uncorrelated case
corresponds to $\lambda = 0$ where $\alpha =1$ and $\beta =
0$. Therefore it is clear that $\kpsi$ is connected to the
Hartree--Fock ground state $\kphi = \ket{g\uparrow g\downarrow}$.
The energy eigenvalues of the exact states are given by
\beq
  \begin{array}{lcl}
  E_0 &=& {{2\,{\cal E} + U - {\lambda \over 2}\,U - 
       {\sqrt{4\,{t^2} + {{\lambda^2 \over 4 }}\,{U^2}}}}} \\
  E_{S_u} &=& 2\,{\cal E} + U  \\
  E_T &=& 2\,{\cal E} + U - \lambda \,U
  \end{array}
\eeq
The excitation energies $\Delta E_i$ are defined as usual by the
difference of the excited state's energy $E_i$ to the ground state
energy $E_0$. The energy related to the excitation into the triplet,
e.~g., is thus given by
\beq \label{exact_E_T}
  \Delta E_T  = E_T - E_0 =  - {\lambda \over 2}\,U +
       {\sqrt{4\,{t^2} + {{\lambda^2 \over 4 }}\,{U^2}}}
\eeq
Note that the expansion of $\Delta E_i$ into a power series in $\lambda$
yields the Rayleigh--Schr{\"o}dinger series of the excitation energies
which has been discussed in a general context in Sec.~\ref{pert_exc}.

\subsection{Results for TDA, RPA, and FOSEP}

We now discuss approximations to the (singlet)
excitation from $\kpsi$ into $\ket{S_u}$ and the (triplet) excitation
into one of the $\ket{T_i}$ states. E.~g.~the triplet excitation into
$\ket{T_{-1}}$ is defined by a particle--hole excitation from the
$\ket{g \downarrow}$ to the $\ket{u \uparrow}$ orbital. The TDA matrix
for this triplet excitation is one dimensional because the excited
state is uncorrelated. Therefore the TDA result for the triplet
excitation energy
\beq
  \Delta E^\tda_T = 2 t -\frac{ \lambda}{2} U 
\eeq
coincides with the first order of Rayleigh--Schr{\"o}dinger
perturbation theory.

The FOSEP matrix $\mat{M}^\sep$ as well as the RPA
matrix, however, have dimension two because of the coupling with the
corresponding hole--particle configuration. The FOSEP matrix for the
triplet excitation is given by
\beq
  \mat{M}^\sep_T =
   \left( \begin{array}{cc}
      2 t -\frac{ \lambda}{2} U & \frac{ \lambda}{2} U \\
      \frac{ \lambda}{2}  U &  -2 t -\frac{ \lambda}{2} U
   \end{array} \right)
\eeq
Its eigenvalues are given by
\beqs
  \Delta E^{\sep (p/u)}_T &= & - {\lambda \over 2}\,U \pm
       {\sqrt{4\,{t^2} + {{\lambda^2 \over 4 }}\,{U^2}}}
\eeqs
As discussed in Sec.~\ref{sec_fosep} we obtain physical and unphysical
eigenvalues of which the latter do not carry any physically relevant
information.  The distinction between the physical and the unphysical
eigenvalue is clear in the present case because $\Delta E^{\sep (p)}_T$
is always non--negative while $\Delta E^{\sep (u)}_T$ is non--positive
for all choices of the parameters. Note that the physical eigenvalue
yields the exact excitation energy $\Delta E^{\sep (p)}_T = \Delta E_T$!

The close relation of the RPA eigenvalue problem to the FOSEP matrix was
discussed in Sec.~\ref{sec_rpa}. The solutions of the RPA equations
for the triplet excitation are given by
\beqs
  \Delta E^{\rpa (p/u)}_T &= & \pm\,{\sqrt{4\,t^2 -
  2\,\lambda\,t \,U}}
\eeqs
Obviously the expression under the square root may become negative for
certain choices of the parameters, in which case the RPA becomes
unstable. 

A plot of the solutions of the FOSEP, the RPA, and the TDA equations
as a function of the perturbation parameter $\lambda$ for a particular
choice of the Hubbard parameters $t$ and $U$ can be found on the top
of Fig.~1. For this choice representing strong interatomic Coulomb
repulsion, the RPA becomes instable. As discussed in
Sec.~\ref{pert_exc} in the framework of a second order analysis, the
RPA lowers the TDA value whereas the FOSEP approximation yields a
higher value for the excitation energy which is correct in the present
case. We have already mentioned that here only the ground state
$\kpsi$ is correlated whereas the excited state $\ket{T_{-1}}$ is
not. This ground state correlation is correctly taken into account by
the FOSEP approximation but not by the RPA. In the present simple
model this goes even beyond second order as can be seen from
Eq.~(\ref{exact_E_T}).

\begin{figure*}
 \begin{picture}(224,303)(0,-5)
  \psfig{figure=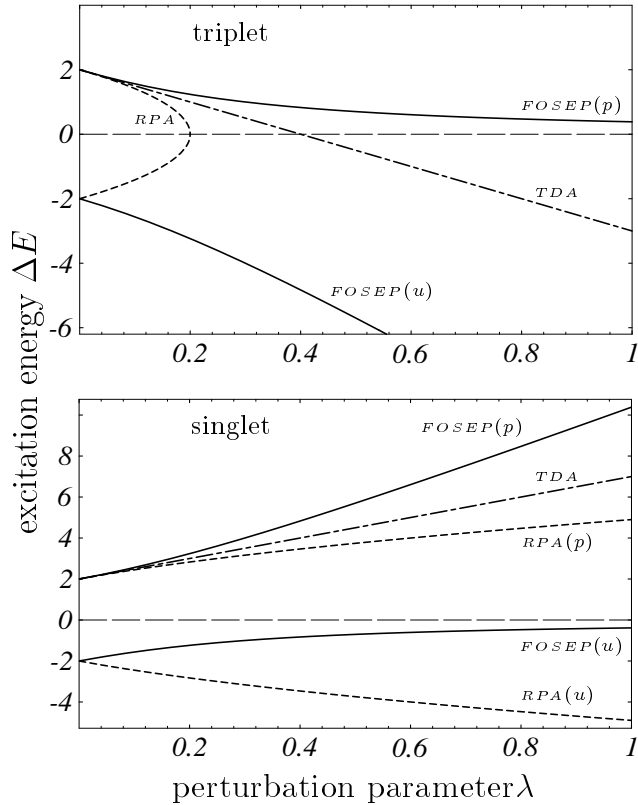,width=224pt,height=298pt,bbllx=0pt,bblly=0pt}
 \end{picture} 
  \caption{The eigenvalues of the FOSEP, RPA, and TDA
  equations for the triplet and the singlet excitation as described in
  the text. The physical FOSEP excitation energies $\Delta E^{\sep
  (p)}$ (labelled as $\sep {\scriptstyle (p)}$) coincide with the
  exact excitation energies. The parameters of the Hubbard model are
  chosen as $t = 1$ and $U=10$. All energies are given in units of
  ${\cal{E}}$. For the triplet excitation, the RPA becomes instable
  yielding complex eigenvalues for $\lambda > 0.2$.}
\end{figure*}

For the singlet excitation, results analogous to the triplet case are
found. Again the FOSEP result $\Delta E^{\sep (p)}_{S_u}$ coincides with
the exact excitation energy $\Delta E_{S_u} = E_{S_u} - E_0$. The TDA
result evaluates to
\beq
  \Delta E^{\tda}_{S_u} = 2 t +\frac{ \lambda}{2} U 
\eeq
while the RPA yields
\beq
  \Delta E^{\rpa (p)}_{S_u} =  {\sqrt{4\,t^2 +
  2\,\lambda\,t \,U}}.
\eeq
A plot of these result with the same choice of parameters as for the
triplet excitation is shown at the bottom of Fig.~1.

The Hubbard parameters used for the plots in Fig.~1 have been chosen
such that the differences between the three first order approximations
are large and become apparent.  When the on--site repulsion term $U$
is decreased in comparison to the hopping parameter $t$, the RPA
becomes stable and the differences between the discussed
approximations diminish. From Eq.~(\ref{aph_und_beth}) it can be seen
that $U$ is the source of ground state correlation in the model. Thus
the present analysis supports the findings of Sec.~\ref{pert_exc} and
leads to the conclusion that among the three considered approximations
for excitation energies only the FOSEP method includes ground state
correlation in a consistent manner.

Of course, it is a special peculiarity of the present simple model
that the FOSEP approximation already yields the exact excitation
energies. Certainly this model may seem inadequate to draw conclusions
on the performance of FOSEP for realistic large finite quantum systems
which constitute the main area of possible applications.  Therefore,
this model should not be understood as a test for large--scale
numerical calculations but rather as the simplest possible system
where the effect of ground state correlation could be investigated
analytically.


\section{Conclusions}

In this paper we have presented a novel approximation scheme for the
calculation of energies and transition moments of many--body systems. It
derives from the first order approximation to the self energy of a
recently proposed extended particle--hole Green's function
\cite{brand_cederbaum_tbp}. The resulting approximations for
excitation energies and transition moments prove consistent in first
order but also higher order terms are present. Starting out from a
Hartree--Fock single--particle description, the FOSEP approximation
yields a matrix eigenvalue problem of the same size as the well known
Random Phase Approximation (RPA). In contrast to the RPA, however,
FOSEP presents a hermitian eigenvalue problem and thus avoids the
instabilities of the RPA. Although the FOSEP approximation has many
properties in common with the RPA, like size consistency and the
invariance with respect to unitary transformations of the
Hartree--Fock virtual orbitals, there are also substantial
differences. We have shown by a perturbation theoretical analysis up
to second order that the FOSEP approximation for the excitation
energies consistently includes part of the ground state correlation
whereas the RPA proves inconsistent in this respect. This statement is
supported by the results of a very simple exactly solvable model. In
the considered model the ground state is correlated whereas the
approximated excited states are not. It turned out that the FOSEP
approximation yields the exact results for the excitation energies
whereas the RPA does worse than the simpler Tamm--Dancoff
approximation (TDA).  We also have addressed the equivalence of length
and velocity form of the transition moments. The exact preservation of
this equivalence is a very peculiar property of the RPA. Within the
FOSEP approximation this equivalence is only preserved in first order
which is consistent for a first order scheme. The second order error
encountered, however, may provide useful in estimating the
applicability of the approximation.

Future calculations on realistic systems still have to provide the
ultimate test for the usefulness of the approximation scheme presented
here. From the present point of investigation the FOSEP method seems
to have excellent prospects to find widespread application like,
e.~g., in clarifying the electronic structure of larger molecules
especially when ground state correlation is important. A possible line
of extending the FOSEP method is to start out from a
multi--configurational self consistent field (MC-SCF) reference state
instead of the Hartree--Fock Slater determinant $\kphi$. This would
allow for an adequate treatment of open--shell or other systems where
strong ground state correlation prohibits the zeroth order description
by a Slater determinant. The first order particle--hole self energy
seems ideally suited for such an extension since the primary matrices
are defined without reference to occupied or virtual Hartree--Fock
orbitals in contrast to the RPA or TDA. Thus the inclusion of
multi--configurational reference states provides a natural extension
of the present theory. 
Straightforward approximations to the static particle--hole self
energy can also be obtained from a ground state description by density
functional theory (DFT). While DFT has been very successful in
predicting ground state properties, the DFT description of excited
states is a vivid and still open field of current interest
\cite{casida95,petersilka96,grimme96,bauernschmitt96}. The static
particle--hole self energy seems well suited for adaption to DFT since
it provides a simple model for excitation properties that only
requires a decent approximate description of one and two particle
densities in the ground state as input as discussed in the
appendix. A direct formulation of the particle--hole self energy in
terms of density functionals, on the other hand, may as well lead to
powerful approximations.  Another open point is the development of
higher order approximations to the particle--hole self energy. This
will allow to increase the accuracy and lift the present restriction
to particle--hole type excitations. A realisation of systematic higher
order approximations could follow the concepts of correlated excited
states and intermediate state representations developed in
Refs.~\cite{mertins1,mertins2} and is left for future work.

\section*{Acknowledgements}

A critical reading of the manuscript and helpful comments by E.~Pahl
and J.~Schirmer are gratefully acknowledged.


\appendix
\section*{}

The general formalism behind the theory of extended two--particle
Green's functions is described thoroughly in 
Ref.~\cite{brand_cederbaum_tbp} where also the definition of the states
$\ky{rs}$, the metric $\hat{\mu}$, and the generalised excitation
energy operator $\cH$ can be found. Here we only want to show one
result of the general theory: The general expression for the primary
block $\mat{\cal H}_{aa}$ of the excitation energy operator matrix is
given by
\beqs \label{H_aa_app}
  \nonumber \lefteqn{
  \left[\mat{\cal H}_{aa}\right]_{rs,r's'}  =  \by{rs} \m \cH
    \ky{r's'}} \\
  \nonumber
  & = &  \bpsi \left[ a^\dagger_s a_r \, , \, \left[ \ham\,,\,
  a^\dagger_{r'} a_{s'} \right]\right] \kpsi \\
  \nonumber
%
%
  & &  + \bpsi \left\{ \left[ \ham\,,\, a^\dagger_{r'}
  \right] \, ,\,  a_r\right\} \kpsi  
       \bpsi a_{s'}  a^\dagger_s \kpsi  \\
  \nonumber
  & & + \bpsi \left\{ a_r\, ,\, a^\dagger_{r'} \right\} \kpsi 
       \bpsi  \left[ \ham\,,\, a_{s'} \right]  a^\dagger_s \kpsi \\
  \nonumber
%
%
  & &  +   \bpsi \left[ \ham\,,\, a^\dagger_{r'} \right]  a_r  \kpsi 
       \bpsi \left\{ a_{s'}\, , \, a^\dagger_s \right\} \kpsi  \\
  & & +   \bpsi a_r a^\dagger_{r'} \kpsi 
       \bpsi \left\{  \left[ \ham\,,\, a_{s'} \right] \, , \,
  a^\dagger_s \right\} \kpsi   .
\eeqs
Here $\kpsi$ denotes the exact ground state of the system.
In general, Eq.~(\ref{H_aa_app}) requires the evaluation of ground
state expectation values of one and 
two--particle operators. This is due to the
particular combination of commutators and anticommutators
and to the fact that the Hamiltonian $\hat{H}$ is
a two--particle operator. But this means that the primary block
$\mat{\cal H}_{aa}$ and therefore the static particle--hole self
energy can be calculated exactly if the general one and two--particle
densities of the ground state are known. 
Approximating the exact ground state $\kpsi$ by a Slater determinant
leads to a factorisation of the two--particle densities and the first
order expression (\ref{fo}) is obtained. Other
approximations for the densities than those obtained by the
Hartree--Hock Slater determinant are of course also possible. 
Density functional theory, on the one hand, or a multi--configurational
self consistent field (MC--SCF) approximation for the ground state
wavefunction, on the other hand, provide interesting alternative approaches.




\begin{thebibliography}{10}

\bibitem{fetter71}
A.~L. {Fetter} and J.~D. Walecka, {\em Quantum Theory of Many--Particle
  Systems}, 1st ed. (McGraw--Hill, New York, 1971).

\bibitem{schirmer82}
J. Schirmer, Phys. Rev. A {\bf 26},  2395  (1982).

\bibitem{ringSchuck}
P. Ring and P. Schuck, {\em The Nuclear Many--Body Problem} (Springer, New
  York, 1980), p.\ 623 ff.

\bibitem{trofimov97}
A.~B. Trofimov and J. Schirmer, Chem. Phys. {\bf 214},  153  (1997).

\bibitem{thouless61}
D.~J. Thouless, Nucl. Phys. {\bf 22},  78  (1961).

\bibitem{rowe68:_eom}
D.~J. Rowe, Rev. Mod. Phys. {\bf 40},  153  (1968).

\bibitem{nielsen80:_soppa}
E.~S. Nielsen, P. J{\o}rgensen, and J. Oddershede, J. Chem. Phys. {\bf 73},
  6238  (1980).

\bibitem{oddershede84}
J. Oddershede, P. J{\o}rgensen, and D.~L. Yeager, Comp. Phys. Rep. {\bf 2},  33
   (1984).

\bibitem{packer96:_soppa}
M.~J. Packer {\it et~al.}, J. Chem. Phys. {\bf 105},  5886  (1996).

\bibitem{rose73:_eom}
J. Rose, T. Shibuya, and V. McKoy, J. Chem. Phys. {\bf 58},  74  (1973).

\bibitem{hansen78}
A.~E. Hansen and T.~D. Bouman, Mol. Phys. {\bf 37},  1713  (1979).

\bibitem{koch95}
H. Koch, O. Christiansen, P. J{\o}rgensen, and J. Olsen, Chem. Phys. Lett. {\bf
  244},  75  (1995).

\bibitem{bauernschmitt96}
R. Bauernschmitt and R. Ahlrichs, Chem. Phys. Lett. {\bf 256},  454  (1996).

\bibitem{brand_cederbaum_tbp}
J. Brand and L.~S. Cederbaum, to be published.

\bibitem{aTarantelli92}
A. Tarantelli and L.~S. Cederbaum, Phys. Rev. A {\bf 45},  2790  (1992).

\bibitem{capuzzi96}
F. Capuzzi and C. Mahaux, Ann. of Phys. (N.Y.) {\bf 245},  147  (1996).

\bibitem{brand96}
J. Brand and L.~S. Cederbaum, Ann. of Phys. (N.Y.) {\bf 252},  276  (1996).

\bibitem{mertins1}
F. Mertins and J. Schirmer, Phys. Rev. A {\bf 53},  2140  (1996).

\bibitem{mertins2}
F. Mertins, J. Schirmer, and A. Tarantelli, Phys. Rev. A {\bf 53},  2153
  (1996).

\bibitem{czyz61}
W. Czy\.z, Acta Phys. Polon. {\bf 20},  737  (1961).

\bibitem{schirmer96}
J. Schirmer and F. Mertins, J. Phys. B {\bf 29},  3559  (1996).

\bibitem{march67}
N.~H. March, W.~H. Young, and S. Sampanthar, {\em The Many--Body Problem in
  Quantum Mechanics} (Cambridge University Press, Cambridge, 1967).

\bibitem{mcweeny89}
R. McWeeny, {\em Methods of Molecular Quantum Mechanics}, 2 ed. (Academic
  Press, London, 1989).

\bibitem{schirmer96:size_consistency}
J. Schirmer and F. Mertins, Int. J. Quant. Chem. {\bf 58},  329  (1996).

\bibitem{szabo82:_moder_quant_chemis}
A. Szabo and N.~S. Ostlund, {\em Modern Quantum Chemistry: Introduction to
  Advanced Electronic Structure Theory} (Macmillan, London, 1982).

\bibitem{oddershede83}
J. Oddershede and J.~R. Sabin, J. Chem. Phys. {\bf 79},  2295  (1983).

\bibitem{harris69}
R.~A. Harris, J. Chem. Phys. {\bf 50},  3947  (1969).

\bibitem{trofimov95}
A.~B. Trofimov and J. Schirmer, J. Phys. B: At. Mol. Opt. Phys. {\bf 28},  2299
   (1995).

\bibitem{jorgensen83}
P. J{\o}rgensen and J. Oddershede, J. Chem. Phys. {\bf 78},  1898  (1983).

\bibitem{ashcroft76:_solid_state_physic}
N.~W. Ashcroft and N.~D. Mermin, {\em Solid State Physics} (Saunders College,
  Philadelphia, 1976), Chap.~32, p.\ 689.

\bibitem{casida95}
M.~E. Casida,  in {\em Recent advanves in density functional methods}, edited
  by D.~P. Chong (World Scientific, Singapore, 1995), Vol.~1.

\bibitem{petersilka96}
M. Petersilka, U.~J. Gossmann, and E.~K.~U. Gross, Phys. Rev. Lett. {\bf 76},
  1212  (1996).

\bibitem{grimme96}
S. Grimme, Chem. Phys. Lett. {\bf 259},  128  (1996).

\end{thebibliography}
\end{document}